\documentclass[showpacs,twocolumn,prl,aps]{revtex4}
\usepackage{graphicx}
\begin{document}

\title{The Hellberg-Mele Jastrow factor as a variational wave function for the one dimensional XXZ model}
\author{Hong-Yu Yang}
\affiliation{Center for Advanced Study, Tsinghua University,
Beijing, 100084, P. R. China}
\author{Tao Li}
\affiliation{Department of Physics, Renmin University of China,
Beijing, 100872, P. R. China}
\date{{\small \today}}

\begin{abstract}
We find the Jastrow factor introduced by Hellberg and Mele in
their study of the one dimensional $t-J$ model provides an
exceedingly good variational description of the one dimensional
XXZ model.
\end{abstract}

\maketitle

The one dimensional systems are genuinely strongly correlated. This
is most clearly indicated by the power law behavior of various
correlation functions, which are termed Luttinger liquid behavior in
general\cite{haldane}. In a Luttinger liquid system, the powers of
the correlation functions, or, the critical exponents of the system,
are continues functions of the model parameter.

The one dimensional XXZ model is typical Luttinger liquid system. In
this short note, we propose a variational description this well
known model. The Hamiltonian of the model reads
\begin{equation}
    \mathrm{H}=-\sum_{i}(\mathrm{S}_{i}^{x}\mathrm{S}_{i+1}^{x}+\mathrm{S}_{i}^{y}\mathrm{S}_{i+1}^{y}
    +\Delta \mathrm{S}_{i}^{y}\mathrm{S}_{i+1}^{y}),
\end{equation}
in which $\Delta$ is the parameter of anisotropy. This model is
exact soluble in terms of the Bethe Ansatz\cite{yangcn}. For
$\Delta<-1$, the system is in the Ising regime in which the ground
state is antiferromagnetically ordered. For $\Delta>1$, the system
phase separates into full polarized regions. For $-1<\Delta<1$,
which is the most interesting case, the system exhibits critical
behavior with continuously varying critical exponents. For example,
the asymptotic behavior of the transverse and the longitudinal spin
correlation function are given by
\begin{equation}
    \langle\mathrm{S}_{i}^{x}\mathrm{S}_{j}^{x}\rangle \sim \frac{A_{x}}{|i-j|^{\eta}}+(-1)^{i-j}\frac{\widetilde{A}_{x}}{|i-j|^{\eta+1/\eta}}
\end{equation}
and
\begin{equation}
    \langle\mathrm{S}_{i}^{z}\mathrm{S}_{j}^{z}\rangle \sim
    \frac{A_{z}}{|i-j|^{1/\eta}}-(-1)^{i-j}\frac{1}{4\pi^{2}\eta|i-j|^{2}},
\end{equation}
in which the critical exponent $\eta$ is given by
\begin{equation}
    \eta=\frac{1}{2}-\frac{1}{\pi}\sin^{-1}\Delta.
\end{equation}

The last equation for the critical exponent $\eta$ is obtained by
combining the result of Bethe Ansatz solution and the effective
field theory based on Abelian Bosonization\cite{luther}. Either the
Bethe Ansatz solution or the effective field theory alone is not
powerful enough to predict the critical exponent. In the Bethe
Ansatz approach, the wave function is so complicated that a direct
calculation of the correlation function is impossible. On the other
hand, while the essence of the critical correlation is well captured
by the effective field theory approach, the parameters in the theory
must be set by hand.

A good variational wave function should capture simultaneously the
short range and the long range correlation of the system studied.
However, for local Hamiltonian, it is difficult to get the correct
long range behavior of the system by optimizing the energy of a
variational wave function. Wave functions very close in energy can
have drastically different long range behavior, provided that the
system is in the critical regime.

We find a wave function originally proposed for the one dimensional
$t-J$ model provides an excellent description of both the short
rang(energy) and the long range correlation of the one dimensional
XXZ model for $-1\leq\Delta\leq1$. The wave function is given by
\begin{equation}
    \Psi_{\mathrm{HM}}(\{
    x_{i}\})=\prod_{i<j}\sin(\frac{\pi(x_{i}-x_{j})}{N})^{\nu},
\end{equation}
in which $x_{i}$ denotes the coordinates of the up spins in the
periodic chain of N sites. This form is first proposed by Hellberg
and Mele to describe the Luttinger liquid behavior of the one
dimensional $t-J$ model\cite{hellberg}. In terms of the $t-J$ model,
this function appears as a Jastrow factor for the residual charge
correlation in front of the well known Gutzwiller projected Fermi
sea wave function. For that purpose, one should reinterpret
$\Psi_{\mathrm{HM}}(\{x_{i}\})$ as a wave function for hard core
Boson, in which $x_{i}$ then denotes the coordinates of the
charges(or holes) in the $t-J$ model.

The function introduced by Hellberg and Mele is also known as the
exact ground state of the Sutherland model with inverse square
interaction in one spatial dimension\cite{sutherland}. The
Sutherland model reads,
\begin{equation}
    \mathrm{H}=-\sum_{i}\frac{\partial^{2}}{\partial
    x^{2}_{i}}+\frac{g\pi^{2}}{L^{2}}\sum_{i<j}\sin^{-2}
    (\frac{\pi(x_{i}-x_{j})}{L}),
\end{equation}
where the last term is a generalized inverse square potential on a
periodic chain of length $L$. It is found that when
\begin{equation}
    \nu=(\sqrt{2g+1}+1)/2,
\end{equation}
$\Psi_{\mathrm{HM}}(\{x_{i}\})$ is the exact ground state of the
Sutherland model.

In our recent work on the variational study of the one dimensional
$t-J$ model, we find the residual charge correlation beyond the
Gutzwiller projected wave function(GWF) of the model should be
described by a XXZ-type effective Hamiltonian\cite{yang}. Combining
this analysis and wave function proposed by Hellberg and Mele, one
quickly realized that the Jastrow factor introduced by them should
also be a good description of the ground state of the one
dimensional XXZ model itself.

The reason for the excellentness of the Hellberg-Mele wave function
for the one dimensional XXZ model can be more directly seen as
follows. At $\Delta=1$, the XXZ model reduces to that of the
isotropic spin chain with ferromagnetic exchange, whose ground state
is the fully polarized state. The wave function for the fully
polarized state is a constant in the Ising basis, which is given
exactly by $\Psi_{\mathrm{HM}}(\{x_{i}\})$ with $\nu=0$. At
$\Delta=0$, the model reduces to the one dimensional XX model, which
is equivalent to the free spinless Fermion model through the
Jordan-Wigner transformation. In this case, the ground state wave
function is given exactly by $\Psi_{\mathrm{HM}}(\{x_{i}\})$ with
$\nu=1$, a Slater determinant for the spinless Fermion. Another
special case is when $\Delta=-1$. In this case, the model reduces to
that of the isotropic spin chain with antiferromagnetic exchange.
Although $\Psi_{\mathrm{HM}}(\{x_{i}\})$ is no longer an exact
ground state for this model, it is well known that the
$\Psi_{\mathrm{HM}}(\{x_{i}\})$ with $\nu=2$, namely a Gutzwiller
projected half filled Fermi sea wave function, provides a exceeding
good variational description for the model considered\cite{gebhard}.
For example, the energy calculated from
$\Psi_{\mathrm{HM}}(\{x_{i}\})$ is about -0.6921 per bond, which
very close to the exact value, $-\ln2$. At the same time, the long
range correlation of the model is also correctly captured by this
wave function, apart from a logarithmic correction which is absent
for $\Delta>-1$.

The ground state energy and other ground state correlation of
$\Psi_{\mathrm{HM}}(\{x_{i}\})$ can be easily calculated by the
Variational Monte Carlo method. Figure 1 shows the ground state
energy calculated from $\Psi_{\mathrm{HM}}(\{x_{i}\})$. We find the
relative error in the ground state energy is less than 0.24 percent(
which is reached at $\Delta=-1$) for all values of $\Delta$.
\begin{figure}[h!]
\includegraphics[width=9cm,angle=0]{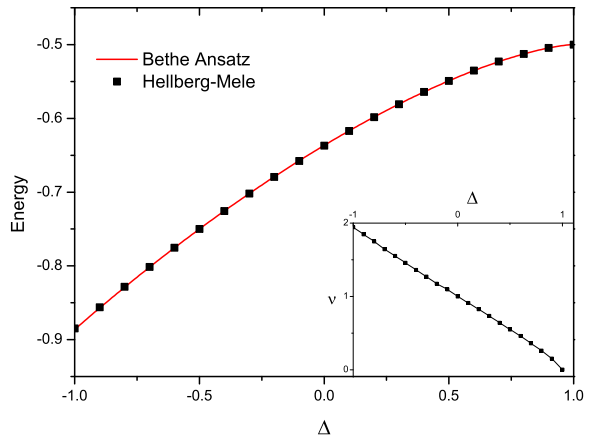}
\caption{Variational ground state energy calculated from
$\Psi_{\mathrm{HM}}(\{x_{i}\})$. The solid line shows the exact
result obtained from Bethe Ansatz solution. The optimized value of
the $\Delta$ is shown in the inset.}
 \label{fig1}
\end{figure}

The Hellberg-Mele-type wave function can not only give precise
estimate for the ground state energy, but also give qualitatively
correct behavior of the critical exponent of the model. In
\cite{kawakami}, it is found that the critical exponent $\eta$ for
$\Psi_{\mathrm{HM}}({x_{i}})$ is simply given by $\frac{\nu}{2}$. In
Figure 2, we compare the exact result for the critical exponent with
that given by the Hellberg-Mele wave function. One find the
variational result agree qualitatively with the exact one. Note for
$\Delta=1$, although the spin rotational symmetric GWF give the
correct value of 1 for the critical exponent $\eta$, one find that
in the variational description in terms of
$\Psi_{\mathrm{HM}}({x_{i}})$ the system chooses to break such
symmetry slightly. We think this failure should be attributed to the
logarithmic coorection at the symmetric point $\Delta=-1$.
\begin{figure}[h!]
\includegraphics[width=9cm,angle=0]{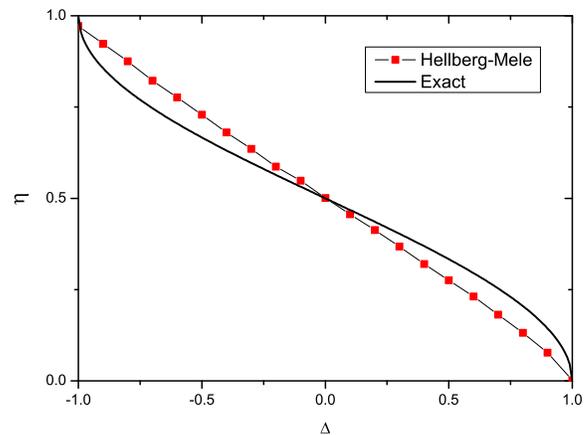}
\caption{Correlation exponent $\eta$ as defined in Eq.(1) as
determined from the Hellberg-Mele wave function. The solid line
denotes the exact result $\eta=1-\frac{1}{\pi}\cos^{-1}\Delta$.}
 \label{fig2}
\end{figure}

Finally, we note that although the Hellberg-Mele wave function is
good approximation for the $\mathrm{S}^{z}_{tot}=0$ sector of the
one dimensional XXZ model, it fails to describe the physics of the
$\mathrm{S}^{z}_{tot}\neq0$ sector of the same model. In
\cite{yang}, this is found to be responsible for the failure of the
Hellberg-Mele-type wave function to describe the Luther-Emery phase
at small electron density and large $J/t$ of the one dimensional
$t-J$ model.

\bigskip This work is supported by NSFC Grant No.90303009.

\end{document}